\documentclass[aps,prb,twocolumn,groupedaddress]{revtex4-1}             
\usepackage{amsmath}
\usepackage{graphicx}
\usepackage{color}
\usepackage{float}

\begin{document}


\title{Anomalous behavior in the phonon dispersion of  the (001) surface of Bi$_2$Te$_3$ determined from helium atom-surface scattering measurements}


\author{Colin Howard, M. El-Batanouny}
\affiliation{Department of Physics, Boston University, Boston, Massachusetts 02215, USA}
\author{R. Sankar, F.C. Chou}
\affiliation{Center of Condensed Matter Sciences, National Taiwan University, Taipei 10617, Taiwan}


\date{\today}

\begin{abstract}
We employ inelastic helium atom-surface scattering to measure the low-energy phonon dispersion along high-symmetry directions on the surface of the topological insulator Bi$_2$Te$_3$. Results indicate that one particular low-frequency branch experiences noticeable mode softening attributable to the interaction between Dirac fermion quasiparticles and phonons on the surface. This mode softening constitutes a renormalization of the real part of the phonon self-energy. We obtain the imaginary part, and hence lifetime information, via a Hilbert transform. In doing so we are able to calculate an average branch specific electron-phonon coupling constant $\langle\lambda_\nu\rangle$ = 1.44.
\end{abstract}


\maketitle

\section{Introduction}
The compound Bi$_2$Te$_3$ is a strong three dimensional topological insulator (TI), a recently discovered class of materials with an insulating bulk electronic structure and protected metallic quasiparticle surface states\cite{Hasan,Qi,Moore1,Fu1,Hasan2}. The surface quasiparticles are found to be Dirac fermions with a single Dirac-cone located in the bulk band-gap and centered at the $\bar\Gamma$-point\cite{Zhang2}. Strong spin-orbit interactions lock the Dirac fermion quasiparticles' (DFQs) spin and wave-vector in a mutually perpendicular configuration, giving the Dirac cone a definite chirality. As a result the surface states are robust against time-reversal invariant perturbations. Consequently, DFQs on the surface cannot backscatter from lattice vacancies, grain boundaries, phonons, etc. into their time reversed counterparts. Despite these constraints, it was recently found that the DFQs strongly couple to surface boson excitations, especially phonon \cite{zhu} and plasmon/spin \cite{Kondo,Raghu} excitations. 
Technical improvements may minimize defects, but phonons are always present. Consequently, DFQ-phonon  interaction should be a dominant scattering mechanism for Dirac fermions on these surfaces at finite temperatures. Hence, the electron-phonon (e-p) interaction is of exceptional importance when assessing the feasibility of promising applications in technologies such as spintronics and quantum computing.


In this paper we present experimental and theoretical studies of the surface phonon dispersions of Bi$_2$Te$_3$ along the high-symmetry directions $\bar\Gamma\bar{\mbox{M}}$ and $\bar\Gamma \bar{\mbox{K}}\bar{\mbox{M}}$. The experimental measurements were carried out with the aid of elastic and inelastic helium atom-surface scattering (HASS). Phonon mode identification was obtained by fitting the results of lattice dynamics slab calculations, based on the pseudocharge model, to experimental data. In order to extract information about the e-p interactions, an expression for the phonon self-energy was derived from a phenomenological model employing the random phase approximation (RPA). The real part of the self-energy was fitted to the experimental results, and, subsequently the imaginary part was obtained with the aid of the Kramers-Kronig transformation. This allowed the determination of mode-dependent e-p coupling, i.e. phonon-branch-specific $\lambda_\nu({\bf q})$, where the subscript $\nu$ indicates the particular branch involved.

The highlights of the study include a strong Kohn anomaly and the absence of long wavelength Rayleigh modes. Moreover, one particular optical surface phonon branch originating at the $\bar\Gamma$ point with $\omega\approx 1.4$ THz experiences significant renormalization owing to interactions with the DFQs. We fitted our experimental results to lattice dynamical calculations based on the pseudo-charge model and found it necessary to introduce a coupling between surface pseudocharge to account for the unique shape of the dispersion. The extracted average value of the branch-specific dimensionless e-p coupling constant is $\langle\lambda_\nu\rangle$ = 1.44. 

In section \ref{sec:expsetup} we discuss the experimental setup and procedures. The formulation of the pseudocharge based lattice dynamical calculations is presented in section \ref{sec:PCM}, while a brief discussion of the RPA machinery is outlined in section \ref{sec:RPA}. Finally we present the results and discussion in section \ref{sec:results}.
 
\section{\label{sec:expsetup} Experimental setup and procedures}

\subsection{Crystal preparation}
Polycrystalline Bi$_2$Te$_3$ compounds were prepared by solid state reaction. High-purity starting elements of Bi (99.999\%, chunks) and Te (99.999\%, shot) were melted in evacuated carbon coated quartz tubes at 550$^\circ$ C for 10 hrs, slowly cooled to 150$^\circ$ C, followed by a rapid quenching in water. The obtained ingots were pulverized into powder, sealed in evacuated quartz tubes (10 cm length and 1.6 cm inner diameter at $\sim10^{-3}$ Torr) after multiple argon gas purging cycles, pretreated at 550$^\circ$ C for 48 hrs in a box furnace and furnace cooled. Single crystals were grown with a vertical Bridgman furnace starting from the pretreated sealed tubes. The temperature profile of the Bridgman furnace used for the whole series was maintained at 350-700$^\circ$ C within a 25 cm region. Initial complete melting was achieved at 700$^\circ$ C for 24 hrs to ensure complete reaction and mixing. A temperature gradient of 0.5$^\circ$ C/cm was programmed around the solidification point near 585$^\circ$ C, and the quartz tube was then slowly lowered into the cooling zone at a rate of ~0.5 mm/h. The grown single crystals were 3 cm long and about 1.6 cm in diameter with good optical quality. They are easy to cleave with crystal planes perpendicular to the hexagonal c-axis.

Grown crystals of Bi$_2$Te$_3$ were cut into wafers approximately 3mm $\times$ 3mm $\times$ 1mm. The wafer was subsequently attached to the OFHC sample holder using UHV compatible conducting epoxy. The sample holder and wafer were baked at 180$^\circ$ C for approximately 1 hour to allow the epoxy to cure. Upon removal an additional layer of epoxy was added to the top of the wafer onto which a cleaving pin was pressed. The entire assembly consisting of sample holder, wafer, and cleaving pin was again baked for roughly 1 hour. After cooling to room temperature the sample assembly was transfered into the UHV chamber and mounted on a sample manipulator equipped with XYZ motions as well as polar and azimuthal rotations. A base pressure of $\sim 10^{-10}$ Torr was maintained in the chamber with the aid of a liquid nitrogen baffle, titanium pump, and turbo molecular pump. The partial pressures of the main contaminants in the chamber (CO, CO$_2$ and H$_2$O) were consistently below  $5.0 \times 10^{-11}$ Torr. The sample was then cleaved \emph{in situ} by knocking the cleaving pin off the wafer. All measurements were performed with the sample surface at room temperature.

\subsection{Experimental Measurements}
Experimental measurements of the surface phonon dispersion were carried out at the Helium atom-surface scattering facility at Boston University. 
Elastic diffraction was used to determine the surface structure and its quality, and to orient the sample surface along desired high-symmetry directions. Inelastic surface scattering methods, based on time of flight (TOF) techniques, were employed to perform measurements of the phonon dispersion. 

\begin{figure}
\includegraphics[scale=0.40]{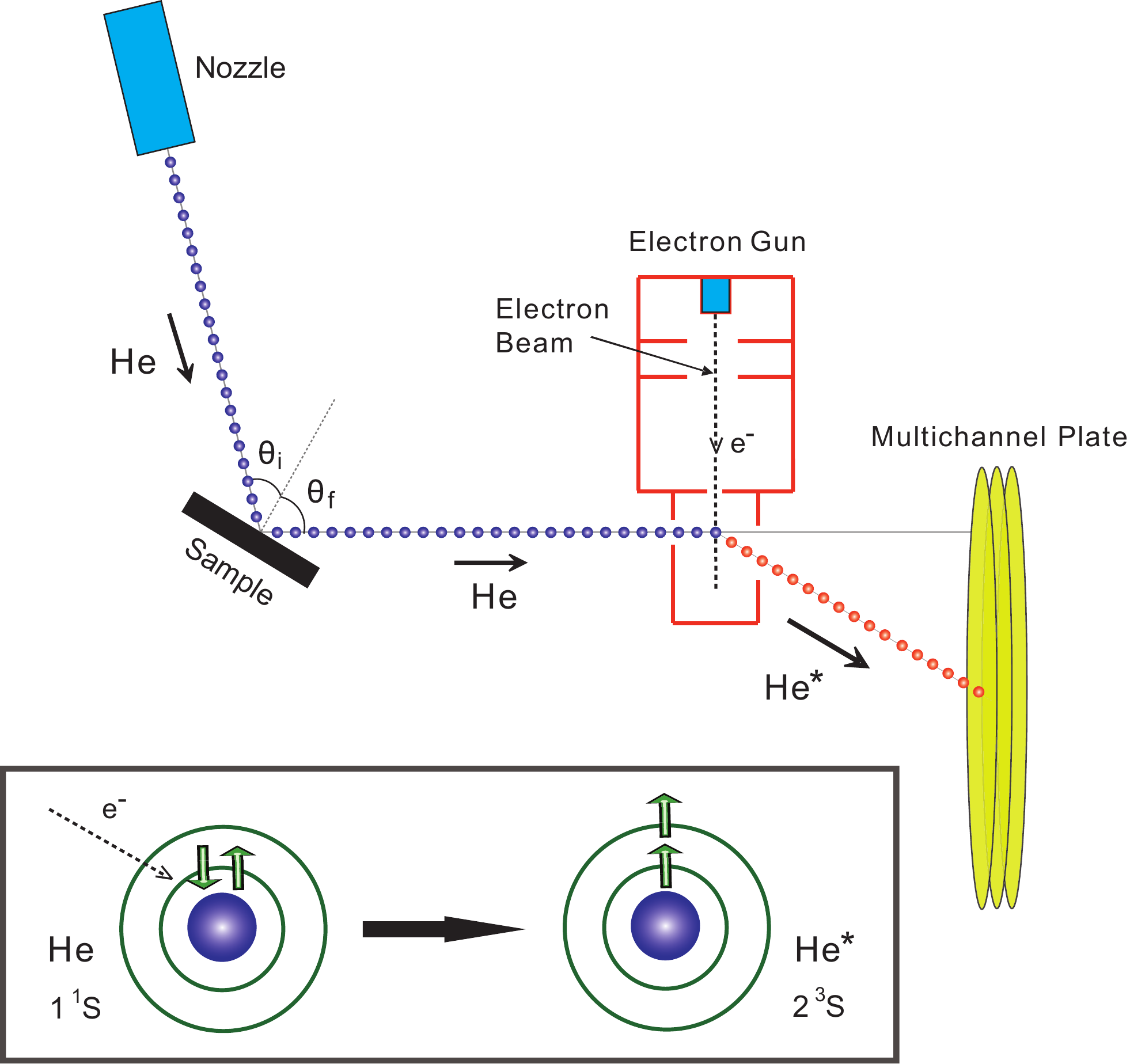}
\caption{Schematic diagram of helium beam detection mechanism. The incident beam energy is controlled by the nozzle temperature and the incident and final angles may be adjusted via the polar controls of the sample manipulator and detector respectively.\label{detector}}
\end{figure}

Detection of the scattered helium and its energy distribution is effected by a  metastable atom velocity analyzer.  As shown in figure \ref{detector}, the detector \cite{Martini} is comprised of an electron gun and a multichannel plate (MCP) electron multiplier. The electron gun generates a well-collimated, monoenergetic electron beam crossing the angle-resolved scattered He beam at right angles. The energy of the electron beam is tuned to excite the He atoms to their first excited metastable state (He*, 2 $^3$S, $10^4$ s lifetime) upon impact. Deexcitation of a He* atom at the surface of the MCP leads to the ejection of an electron in a manner similar to Auger emission, which generates an electron cascade that is then collected by the anode of the multiplier. By electronically pulsing the electron gun, a gate function is created for TOF measurements in the inelastic HASS mode. The details of the detection scheme are given in Ref 10.

\begin{figure}
\includegraphics[scale=0.60]{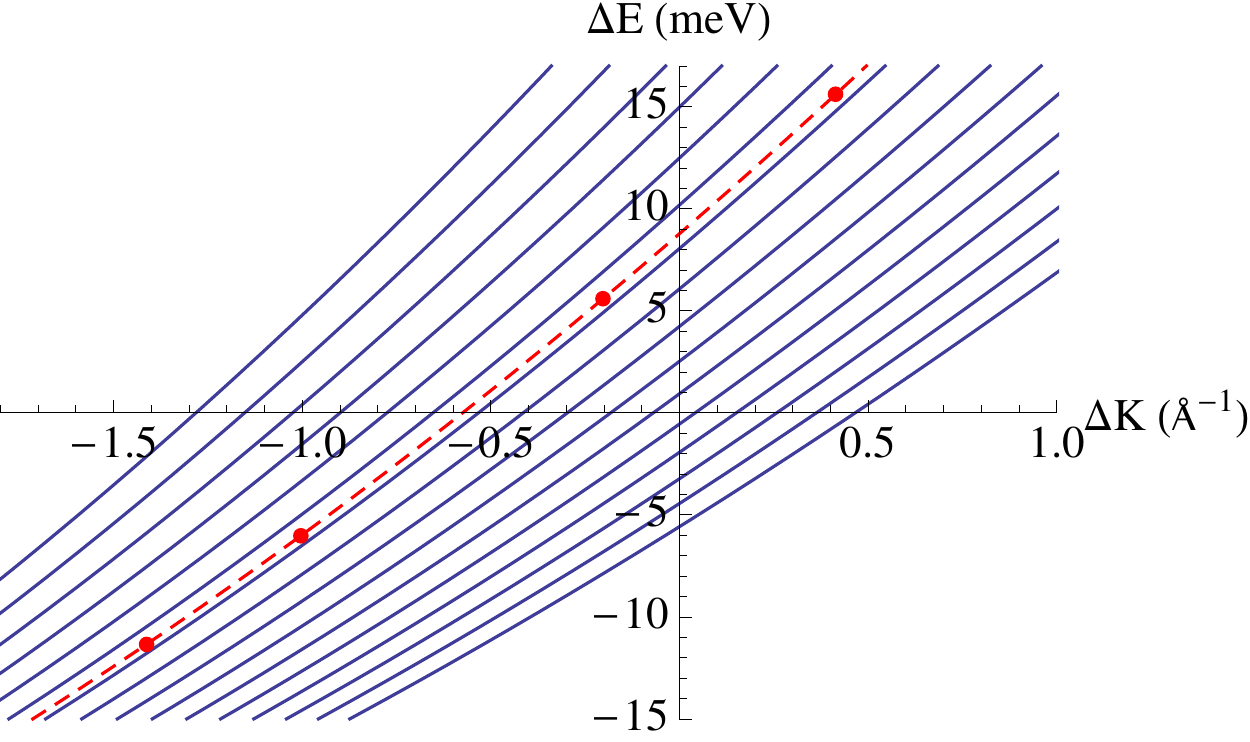}
\caption{Scan curves showing the kinematically allowed scattering events for $E_i$ = $43.09$ meV and $\theta_i$ = 45.62$^\circ$. Each blue parabola corresponds to a unique $\theta_f$, which are plotted here in 1$^\circ$ increments from 35$^\circ$ to 50$^\circ$. The red parabola and data points correspond to the phonon events shown in the third panel of figure \ref{TOF}. \label{SC}}. 
\end{figure}

By writing the He-atom wave-vector as ${\bf k}=({\bf K},k_z)$, where ${\bf K}$ is the component parallel to the surface, conservation of momentum and energy for the `in-scattering-plane' geometry can be expressed as
\begin{align}\label{1a}
\Delta K\,&=\,|{\bf G}+{\bf Q}|\,=\,k_{f}\sin\theta_{f}-k_{i}\sin\theta_{i}\\\label{2a}
\Delta E\,&=\,\hbar\omega({\bf Q})\,=\,\frac{\hbar^2k_f^2}{2M}-\frac{\hbar^2k_i^2}{2M}
\end{align}
where subscripts $i$ and $f$ denote incident and scattered beams, respectively, and $\Delta K$ is the momentum transfer parallel
to the surface. ${\bf G}$ is a surface reciprocal-lattice vector, ${\bf Q}$ is the surface phonon wave-vector, $\hbar\omega({\bf
Q})$  is the corresponding surface phonon energy and $M$ is the mass of a He atom. By eliminating $k_f$ from the above equations, one obtains the so-called {\sl scan curve} relations which are the locus of all the allowed $\Delta K$ and $\Delta E$ as dictated by the geometry and the conservation relations,
\begin{equation} 
\Delta E\,=\,E_i\,\left[\left(\frac{\sin\theta_i + \Delta K/k_i}{\sin\theta_f}\right)^2-1\right]
\end{equation}
where $E_i=\hbar^2k_i^2/2M$. Note that the energy exchange $\Delta E$ can be both positive and negative, corresponding to phonon annihilation and creation, respectively. A sample set of scan curves for fixed $E_i$ and $\theta_i$ but variable $\theta_f$ may be seen in figure \ref{SC}. The intersections of these scan curves with the phonon dispersion curves define the kinematically allowed inelastic events for a fixed geometric arrangement. Thus, the geometric scattering configurations for scanning a particular region of $\Delta E$ versus $\Delta K$ can be pre-selected. By systematically changing $E_i,\,\theta_i$, and $\theta_f$, the entire set of dispersion curves along a particular direction in the Brillouin zone can be accessed.

\section{\label{sec:PCM}Pseudocharge Phonon Model}
\subsection{Crystal Structure}
\begin{figure}
\includegraphics[scale=0.45]{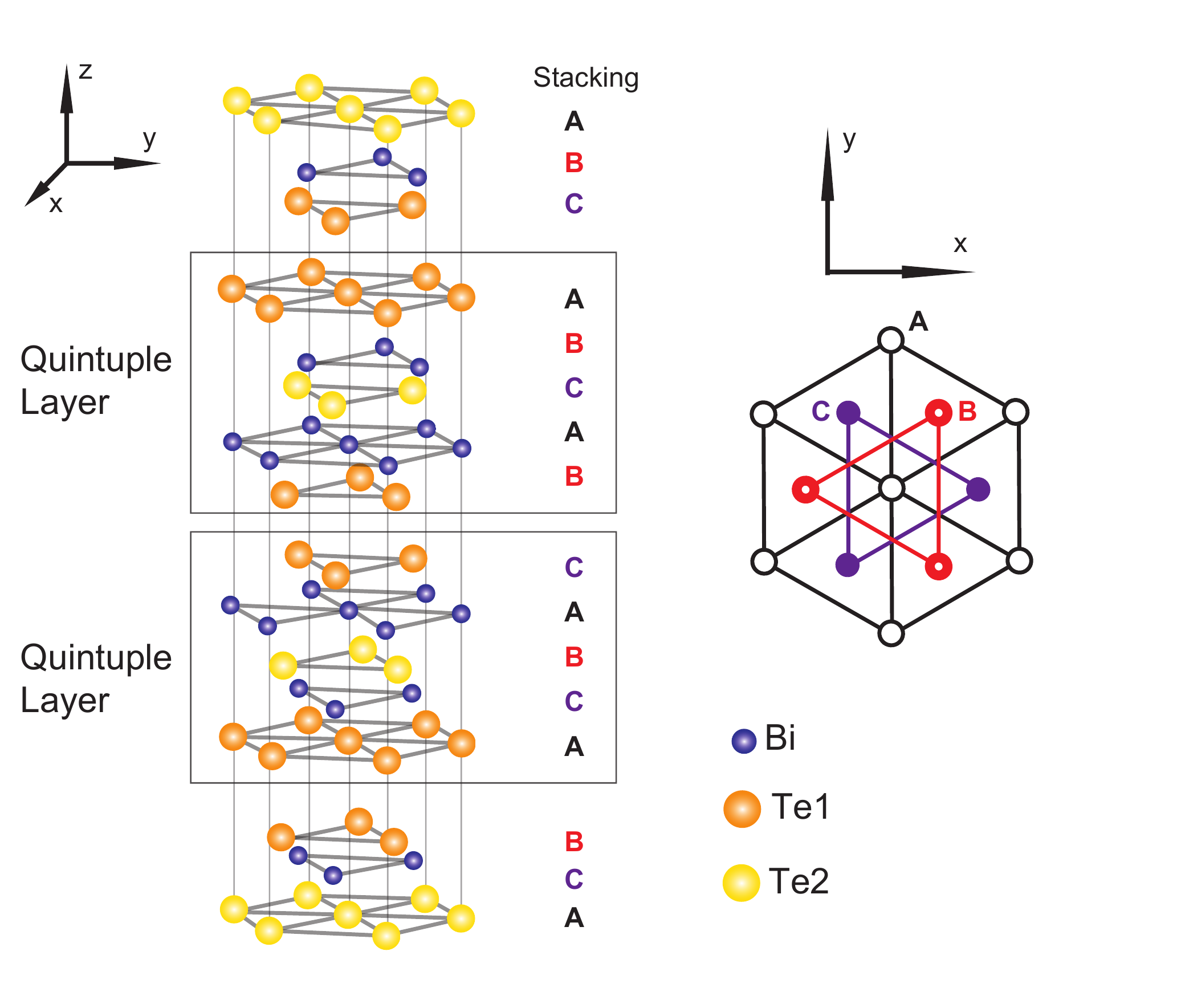}
\caption{Hexagonal unit cell of the Bi$_2$Te$_3$ crystal comprised of three QLs and belonging to the space group $R\bar{3}m$. Note that the Te2 layer within each QL is a center of inversion symmetry. \label{realstructure}}. 
\end{figure}

The Bi$_2$Te$_3$ primitive cell is shown in figure \ref{realstructure}. Alternating hexagonal monatomic crystal planes stack in ABC order. Units of Te-Bi-Te-Bi-Te form quintuple layers (QLs): bonding between atomic planes within a QL is covalent whereas bonding between adjacent QLs is predominantly of the Van der Waals type. The primitive cell contains 3 QLs. The crystal structure belongs to the space group $R\bar{3}m$. The point group contains a binary axis (with twofold rotation symmetry), a bisectrix axis (appearing in the reflection plane) and a trigonal axis (with threefold rotation symmetry). 
The primitive translation vectors in the hexagonal basis are 
\begin{equation}
{\bf t}_1 = a\bigg(\frac{\sqrt{3}}{2},-\frac{1}{2},0\bigg), \ {\bf t}_2 = a(0,1,0), \  {\bf t}_3 = c(0,0,1)
\end{equation}
where $a$ and $c$ are lattice constants of the hexagonal cell. 

\begin{figure}
\includegraphics[scale=0.45]{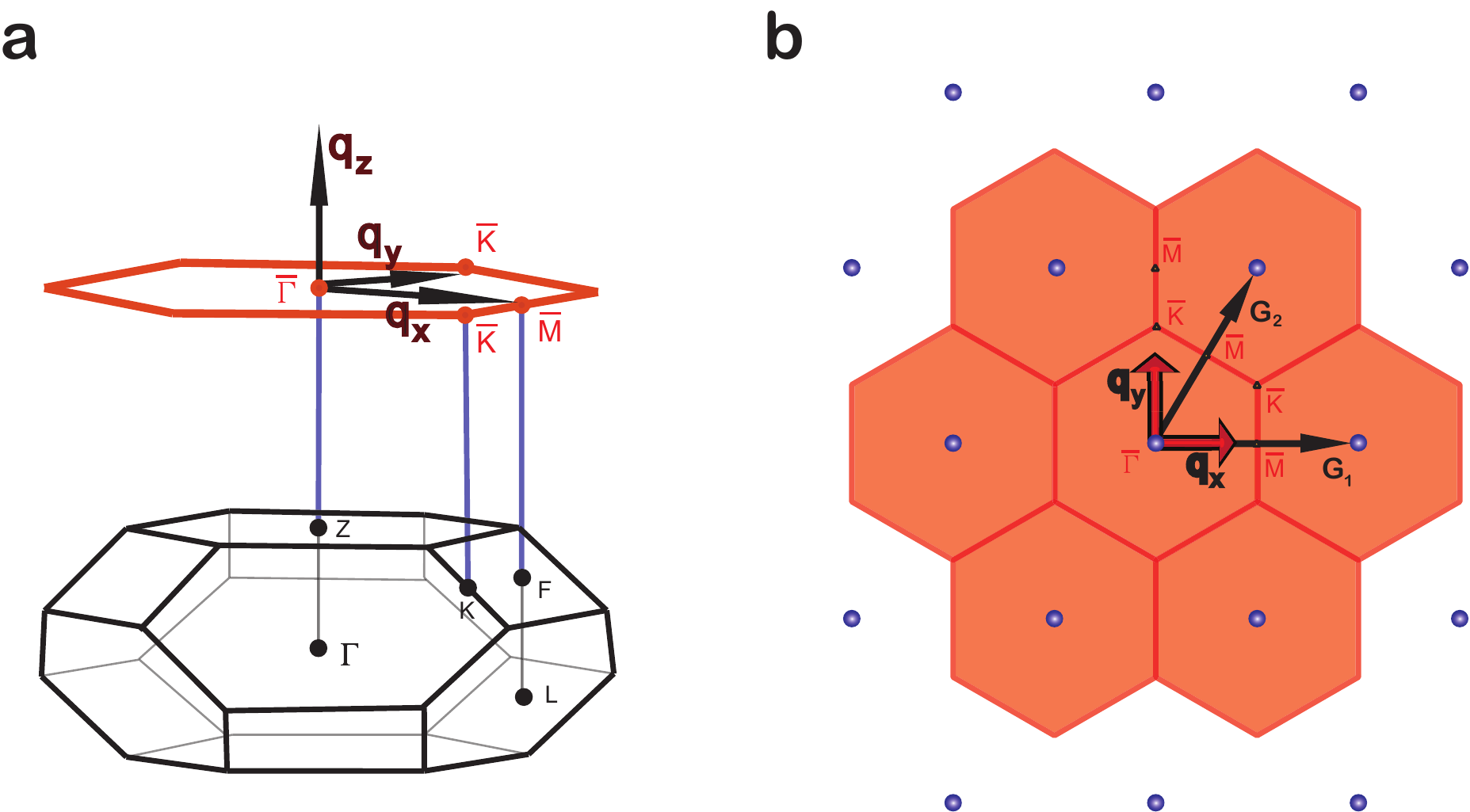}
\caption{Bulk and surface reciprocal space structure of Bi$_2$Te$_3$. Panel {\bf a} shows the bulk Brillouin zone and high-symmetry points. The surface Brillouin zone is represented as a projection along ${\bf q_z}$. The extended surface reciprocal lattice is shown in {\bf b} depicting high-symmetry points $\bar\Gamma, \bar{\mbox{M}}$, and $\bar{\mbox{K}}$ as well as reciprocal lattice vectors ${\bf G}_1$ and ${\bf G}_2$ . \label{recipstructure}}. 
\end{figure}

The corresponding reciprocal lattice vectors are 
\begin{eqnarray}
{\bf G}_1 &=&  \frac{2\pi}{a}\bigg(\frac{2}{\sqrt{3}},0,0\bigg), \  {\bf G}_2 =  \frac{2\pi}{a}\bigg(\frac{1}{\sqrt{3}},1,0\bigg) \nonumber \\
{\bf G}_3 &=& \frac{2\pi}{c}(0,0,1)
\end{eqnarray}
The reciprocal space structure of Bi$_2$Te$_3$ is shown in figure \ref{recipstructure}. 

\subsection{Equations of motion}
In order to identify the character and symmetry of the measured inelastic events, we employ empirical lattice dynamics slab calculations, based on the pseudocharge model (PCM)\cite{Jayanthi, Benedek, Kaden}. With this procedure we characterize and further substantiate our measurements of the phonon dispersion of Bi$_2$Te$_3$. Here we review some of its basic characteristics and tabulate the parameters used in our realization of the model. 

To begin, we expand the electron density $n_l$ within each primitive cell $l$ in terms of symmetry-adapted multipole components around selected Wyckoff symmetry points ${\bf R}_{lj}$. 
\begin{equation}
n_l({\bf r}) = \sum_{j \Gamma k}c_{\Gamma k}Y_{\Gamma k}({\bf r}-{\bf R}_{lj})
\end{equation}
where $\Gamma$ denotes an irreducible representation (irrep) of the Wyckoff symmetry point-group and $k$ indexes its rows; $Y_{\Gamma k}$ is a symmetry-adapted harmonic function. The expansion coefficients $c_{\Gamma k}$ are separated into static and dynamic components, the latter being treated as bona-fide time-dependent dynamical variables,  
\begin{equation}
c_{\Gamma k}(t) = c_{\Gamma k}^{(0)} + \Delta c_{\Gamma k}(t)
\end{equation}

We may write the Lagrangian of the combined pseudocharge-ion system as 
\begin{widetext}
\begin{equation}
{\mathcal L} = \frac{1}{2}\bigg(\sum_{l\kappa\alpha}M_\kappa \dot u_\alpha^2(l\kappa) + \sum_{\substack{\Gamma k\\ lj}}m_{\Gamma} \Delta \dot c_{\Gamma k}^2(lj)\bigg) - \bigg(\frac{1}{2}\Big[{\bf u\cdot\Phi\cdot u + (u\cdot T \cdot\Delta c + h.c.) + \Delta c \cdot H\cdot \Delta c}\Big] \bigg)
\end{equation}
\end{widetext}
The kinetic term contains contributions from both the ions and pseudocharge (PC) where $u_\alpha(l \kappa)$ is the displacement in the $\alpha$ direction of the ion at site $\kappa$ in unit cell $l$; $M$ is the ionic mass, and $m_\Gamma$ is an effective PC mass that will be set to zero upon invoking the adiabatic approximation. The potential energy is expanded in a Taylor series to second order in the ion (PC) displacements (deformations). ${\bf \Phi}$, ${\bf T}$, and ${\bf H}$ are empirical force-constant matrices representing ion-ion, ion-PC, and PC-PC interactions, respectively. 

We obtain the Euler-Lagrange equations of motion for the ions and PCs in the standard way,
\begin{align}
M_\kappa \ddot u_\alpha(l\kappa) =& -\sum_{l^\prime\kappa^\prime\beta}\Phi_{\alpha\beta}\Big(\substack{l \ l^\prime \\ \kappa \ \kappa^\prime}\Big)u_\beta(l^\prime\kappa^\prime) \notag \\ &- \sum_{\substack{l^\prime j \\ \Gamma k}} T_{\substack{\alpha \\ \Gamma k}}\Big(\substack{l \ l^\prime \\ \kappa \ j}\Big)\Delta c_{\Gamma k}(l^\prime j)\label{eomion} \\
m_\Gamma \Delta \ddot c_{\Gamma k}(lj) =& -\sum_{l^\prime\kappa^\prime\alpha}T_{\substack{\Gamma k \\ \alpha}}\Big(\substack{l \ l^\prime \\ j \ \kappa^\prime}\Big)u_\alpha(l^\prime\kappa^\prime) \notag \\ &- \sum_{l^\prime j^\prime}\;H_{\Gamma k}\Big(\substack{l \ l^\prime \\ j \ j^\prime}\Big)\Delta c_{\Gamma k}(l^\prime j^\prime) \label{eompc}
\end{align} 
noting that only PCs belonging to the same irrep and same row can couple.

\subsection{Adiatbatic approximation, ionic self-terms, and PC self-terms}
The entries of the matrix ${\bf \Phi}$ are expressed in terms of empirical parameters. However, the diagonal, self-term, elements that determine how the displacement of a particular ion affects its own motion must be treated carefully. We approach this case by first invoking the adiabatic approximation in which we set $m_\Gamma$ = 0. This is equivalent to saying that the electronic response to lattice deformations is instantaneous. We then have
\begin{equation}
\Delta {\bf c} = -{\bf H}^{-1}{\bf T}^T{\bf u} 
\label{PCdisp}
\end{equation}

We now note that the crystal is invariant under an arbitrary rigid displacement ${\bf u_0}$. In this scenario we have 
\begin{equation}
\Delta {\bf c} = -{\bf H}^{-1}{\bf T}^T{\bf u_0}. 
\end{equation}
Substituting for $\Delta {\bf c}$ in (\ref{eomion}), we obtain
\begin{equation}
{\bf 0} =  -{\bf \Phi u_0} + {\bf TH}^{-1}{\bf T}^T{\bf u_0} \label{inv}
\end{equation}
where {\bf 0} is a null column matrix of length $3N$ where $N$ is the number of ions.  Equation (\ref{inv}) is actually satisfied for each $3\times3$ ionic matrix and any arbitrary displacement, provided it is uniform. Thus, we rearrange and separate the self-term from the rest of the sum to obtain

\begin{align}
 {\bf \Phi} \Big(\substack{l \ l \\ \kappa \kappa}\Big)= &-\sum^\prime_{l^\prime\kappa^\prime}{\bf \Phi}\Big(\substack{l \ l^\prime \\ \kappa \kappa^\prime}\Big)\notag\\& + \sum_{\substack{l^\prime j j^\prime \\ \Gamma k }}\;{\bf T}_{\Gamma k}\Big(\substack{l \ l^\prime \\ \kappa j}\Big){\bf H}^{-1}_{\Gamma k}\Big(\substack{l \ l^\prime \\ j j^\prime}\Big){\bf T}_{\Gamma k}^T\Big(\substack{l \ l^\prime \\ \kappa j}\Big)\label{eq:needslabel}
\end{align} 
where the prime on the first sum indicates that the self-term is excluded. 

We proceed in a similar manner to calculate the diagonal elements of the matrix {\bf H}. We again employ translational invariance but this time explicitly choose our rigid displacement to be in the $z$-direction for clarity. We can rewrite equation (\ref{eompc}) as  
\begin{equation}
0 = -\sum_{l^\prime\kappa^\prime}T_{\substack{\Gamma k \\ z}}\Big(\substack{l \ l^\prime \\ j \ \kappa^\prime}\Big)u_{z}(l^\prime\kappa^\prime) - \sum_{l^\prime j^\prime}H_{\Gamma k}\Big(\substack{l \ l^\prime \\ j \ j^\prime}\Big)\Delta c_{\Gamma k}(l^\prime j^\prime)
\label{invpc}
\end{equation} 
We separate the PC self-term from the rest of the sum and rearrange to obtain
\begin{align}
H_{\Gamma k}\Big(\substack{l \ l \\ j \ j}\Big) =& \frac{-1}{\Delta c_{\Gamma k}(l,j)}\bigg(\sum_{l^\prime \kappa^\prime}T_z\Big(\substack{l \ l^\prime \\ j \ \kappa^\prime}\Big)u_z(l^\prime,\kappa^\prime)\notag\\&\hspace{0.5in} + \sum_{l^\prime j^\prime}^\prime H\Big(\substack{l \ l^\prime \\ j \ j^\prime}\Big)\Delta c_{\Gamma k}(l^\prime,j^\prime)\bigg) 
\end{align}
For a rigid displacement of the entire crystal (including the PC) in the $z$-direction we have $\Delta c_{\Gamma k}$ = $u_z$ = $u_0$ for all $l,\kappa,j$.  The PC self-term then takes the form
\begin{equation}
H_{\Gamma k}\Big(\substack{l \ l \\ j \ j}\Big) = -\sum_{l^\prime \kappa^\prime}T_z\Big(\substack{l \ l^\prime \\ j \ \kappa^\prime}\Big) - \sum_{l^\prime j^\prime}^\prime H_{\Gamma k}\Big(\substack{l \ l^\prime \\ j \ j^\prime}\Big)
\label{PCself} 
\end{equation}

With the ionic and PC self-terms fixed we are now in a position to calculate the phonon frequencies. Fourier transforming equation (\ref{eomion}), scaling the force-constant matrices by the ionic masses, and employing equation (\ref{PCdisp}) yields
\begin{eqnarray}
&\omega^2({\bf q}){\bf u}({\bf q}) = {\cal D}({\bf q}){\bf u}({\bf q}) \nonumber \\ & \\
&{\cal D}({\bf q}) = {\bf \Phi}({\bf q}) - {\bf T({\bf q})H}^{-1}({\bf q}){\bf T}^\dagger({\bf q}) \nonumber 
\end{eqnarray}
Thus, to determine the phonon frequencies at a particular wave-vector one needs only construct the dynamical matrix ${\cal D({\bf q})}$ and find its eigenvalues. 

\subsection{Bulk and Surface Parameters}
\begin{figure}
\includegraphics[scale=0.40]{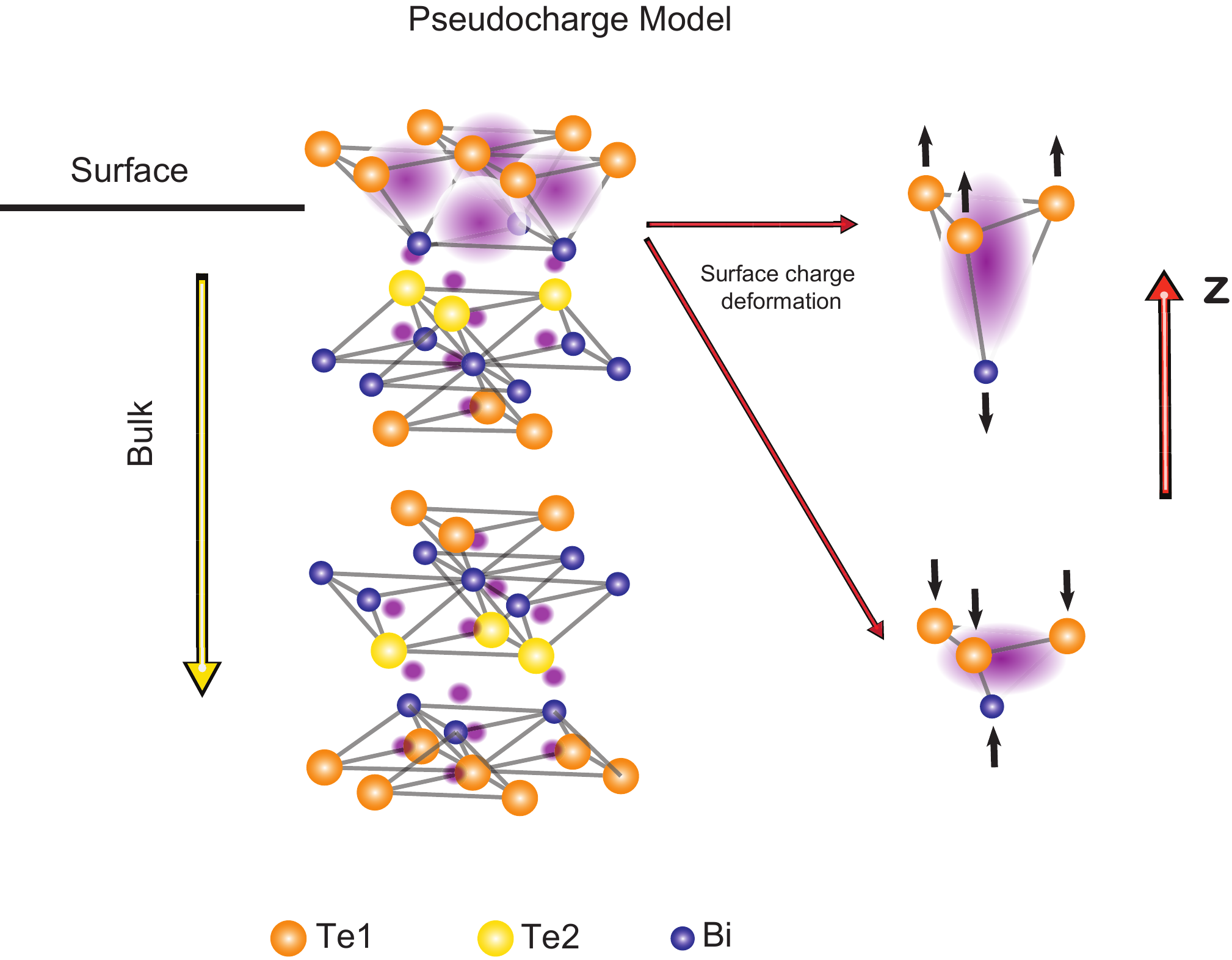}
\caption{Schematic diagram of ion and PC locations in the PCM. The figure indicates that the surface PC are more spread and easier to deform than their bulk counterparts. The figures to the right show the dipolar PC deformation associated with lattice distortions along the $z$ direction.\label{PCSchem}}. 
\end{figure}

In our computational model first the ionic positions are populated. As for the PCs, the `$c$' Wyckoff positions of the $R\bar{3}m$ space group, with $C_\text{3v}$ point-group symmetry, were the most appropriate to use as centers of PC symmetry-adapted multipole expansion. They are identified as having coordinates $(0,0,\pm z)$ that define the vertical axes of the tetrahedral pyramids shown in figure \ref{PCSchem}. The pyramid centers were chosen as PC expansion points. $C_\text{3v}$ has irreps $A_1\,(\text{with dipolar symmetry-adapted harmonic }z)$ and $E\,(\text{with dipolar symmetry-adapted harmonics }x,y)$. In order to minimize the number of empirical constants employed in the bulk calculations, we opted to include only the $A_1$ symmetry-adapted fluctuations as depicted in figure \ref{PCSchem},
which also shows the ion and PC locations in the PCM.

In the insulating bulk we do not include interactions between PCs, rendering {\bf H} diagonal and constrained by equation (\ref{PCself}). However, we introduce two force-constant parameters $T_z^1$ and $T_z^2$ to account for the ion-PC coupling in pyramids involving Te1-Bi and Te2-Bi, respectively. We use central ion-ion interaction potentials $v(r)$ with force-constant matrix elements of the form
\begin{equation}
\Phi_{\alpha\beta} = A\frac{x_\alpha x_\beta}{r_0^2} - B\bigg(\frac{x_\alpha x_\beta}{r_0^3} - \frac{1}{r_0}\delta_{\alpha\beta}\bigg)
\end{equation}   
The parameters $A$ and $B$ are related to the ion potential via 
\begin{equation}
A = \frac{\partial^2v}{\partial r^2}\Big|_{r=r_0} \hspace{5 mm} B = \frac{\partial v}{\partial r} \Big|_{r=r_0}
\end{equation}
where $r_0$ is the equilibrium bond length. We consider only nearest neighbor interactions for Bi-Te couplings. However, because of the large size of Bi atoms we admit coupling to their nearest in-plane (intralayer) Bi neighbors, which are actually second-neighbors, as well as to the nearest Bi atom in the other Bi layer within a single QL (interlayer). In order to determine the force-constant and PC parameters we fit our bulk phonon calculation to available Raman, IR, and inelastic neutron spectroscopy data\cite{Richter,Kullman1,Kullman2,Wagner}. A summary of the parameters used and their values are given in table \ref{bulkpars}.
\begin{table} [h]
\begin{center}
\caption{Bulk parameters for the lattice dynamical calculation based on the PCM.}
\label{bulkpars}
\begin{tabular}{|l| l| l||l l|}
\hline\hline
\multicolumn{3}{|c||}{Ion-ion interaction} & \multicolumn{2}{c|}{Ion-PC interaction} \\[3pt]
\hline
Bond & A (N/m) & B (N) & Position & Value (N/m) \\[3pt]
\hline
Te1-Te1& 0.187 & 0.0187 & $T_z^1$ (Te1-Bi)&0.35 \\[3pt]
Te1-Bi &0.99 &0.099 &$T_z^2$ (Te2-Bi)&0.4 \\[3pt]
Te2-Bi &0.2 &0.02& &\\[3pt]
Bi-Bi (intra)&0.2 &0.02& &\\[3pt]
Bi-Bi (inter)&0.2 &0.02& &\\[3pt]
\hline
\end{tabular}
\end{center}
\end{table}

\begin{figure}
\includegraphics[scale=0.6]{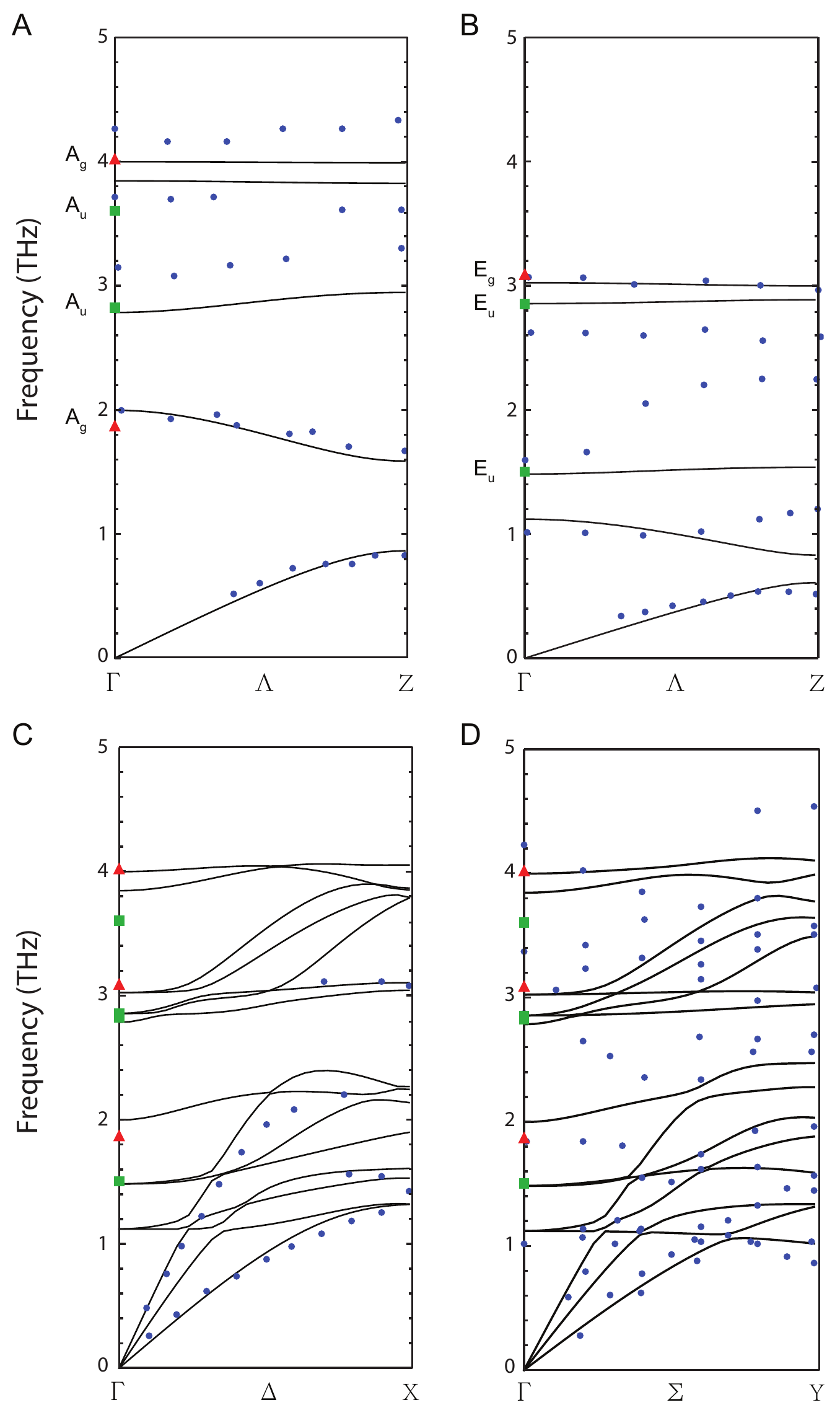} 
\caption{Calculated bulk dispersion curves along the high-symmetry directions $\Lambda$ (A-B), $\Delta$ (C), and $\Sigma$ (D). The $C_{3v}$ symmetry of the $\Lambda$-direction allows one to project out purely longitudinal $A$ modes and doubly degenerate, transverse $E$ modes. Modes along the $\Delta$ and $\Sigma$ directions have mixed polarization. The calculated dispersions were fit to available Raman (red triangles), IR (green squares), and inelastic neutron scattering (blue circles) data. \label{bulkdisp}}
\end{figure}

The best fit to the available bulk data is shown in figures \ref{bulkdisp}A-\ref{bulkdisp}D. 
Dispersions are presented along three high-symmetry directions $\Lambda$ ($\Gamma$-$Z$), $\Delta$ ($\Gamma$-$X$), and $\Sigma$  ($\Gamma$-$Y$). Raman, IR, and neutron data are depicted as red triangles, green squares, and blue circles respectively. Although the calculation agrees well with experimental data at the $\Gamma$ point, we note that there is some discrepancy between our calculation and the neutron data, especially at high phonon energies. Previous studies \cite{Kullman1} have noted this difficulty in replicating experimental data off the $\Gamma$ point without long range Coulomb interactions. Nonetheless the low-energy modes, which we are primarily concerned with, fit quite well. Moreover, the Coulomb interactions in the surface layers are effectively screened due to the presence of DFQs.  

In calculating the surface phonon dispersions we employed a slab geometry containing 30 QLs. In order to obtain the best fit to the experimental data the following changes and additions to the bulk parameter values were made:
\begin{enumerate}
\item
The surface Te1-Bi force-constant parameter was reduced to roughly 42\% of its bulk value to account for the reduced bonding and the emergence of metallic electrons.
\item 
A new planar force-constant parameter involving intralayer surface Te1-Te1 bonds was introduced.
\item
Symmetry-adapted $x/y$ deformations of the PC in the surface and subsurface pyramids, which form a basis of the doubly-degenerate irrep $E$, were introduced to account for the delocalized nature of the DFQs. These are effected via new parameters $T_{xy}^S$ and $\tilde{T}_{xy}^S$, respectively. In addition, $T_z^S$ was reduced from its bulk value to account for the extra screening provided by the DFQ surface states. 
\item 
A momentum dependent coupling $H_q$ between dipolar $z$ deformations of neighboring surface PC was introduced to account for interactions among the DFQs. 
\end{enumerate}
The surface parameters are summarized in table \ref{surfpars}. Calculated dispersions for the slab geometry are presented in section \ref{sec:results}.
\begin{table}
\caption{Modified surface parameters for the lattice dynamical calculation.}
\label{surfpars}
\begin{tabular}{|l| l| l||l l|}
\hline\hline
\multicolumn{3}{|c||}{Surface ion-ion interaction} & \multicolumn{2}{c|}{Surface ion-PC interaction} \\[3pt]
\hline
Bond & A (N/m) & B (N) & Position & Value (N/m) \\[3pt]
\hline
Te1-Bi & 0.42&0.042 &$T_z^S$ (Te1-Bi)&0.24\\[3pt]
Te1-Te1 (intra) & 0.25 &0.025&$T_{xy}^S$ (Te1-Bi)&0.15\\[3pt]
 & & &$\tilde{T}_{xy}^{S}$ (Te2-Bi) &0.4\\[3pt]
\hline
\hline
\multicolumn{5}{|c|}{Surface PC-PC interaction}\\[3pt]
\hline
\multicolumn{5}{|c|}{}\\
\multicolumn{5}{|c|}{$H_q = H_0(1+\frac{q^2}{a}e^{-q^2/b})$, \ $H_0 = -0.0782$, \ $a=0.0034$, \ $b =0.0075$}\\[3pt]
\hline
\end{tabular}
\end{table}

\section{\label{sec:RPA}Calculation of e-p coupling constant in the Random Phase Approximation}
In this section we describe the phenomenological model-fitting approach to the experimentally measured dispersion of the optical phonon branch that exhibits strong e-p renormalization, and the procedure followed to extract the corresponding e-p coupling function $\lambda_\nu({\bf q})$. The construction of the model is carried out with the aid of the RPA.

We start by defining the noninteracting, or free, surface phonons Hamiltonian in second-quantized form
\begin{equation}
\label{eq:phonon Hamiltonian}
{\cal H}_{\rm{ph}}\,=\,\sum_{{\bf q},\nu}\,\hbar\omega^{(0)}_{{\bf q},\nu}\,\left( b^{\dagger}_{{\bf q},\nu}\,b_{{\bf q},\nu} + \frac{1}{2} \right)
\end{equation}
where $b^{\dagger}_{{\bf q},\nu}$ is the creation operator of a phonon of bare frequency $\omega^{(0)}_{{\bf q},\nu}$ and branch index $\nu$. The free phonon Matsubara Green's function of the $({\bf q},\nu)$ mode is defined as
$$\mathcal{D}_\nu^{(0)}({\bf q}, \mathrm{i}\omega_n)\,=\,\frac{2\left(\hbar\omega^{(0)}_{{\bf q},\nu}\right)}{(\mathrm{i}\omega_n)^{2}-\left(\hbar\omega^{(0)}_{{\bf q},\nu}\right)^2}$$ where $\mathrm{i}\omega_n$ is the Matsubara frequency.

The electronic surface states of Bi$_2$Te$_3$ form a two dimensional Dirac metal, whose low-energy physics is well described by the Hamiltonian
\begin{equation}
\label{DH}
{\cal H}_{\rm {el}}\,=\,\sum_{\bf{k}}\;\psi^{\dagger}_{\bf{k}}\,\left[\hbar\, \mathrm{v}_{F}\,\hat{\mathbf{z}}\cdot(\mathbf{k}\boldsymbol{\times\sigma})-\mu\right]
\,\psi_{\bf{k}}
\end{equation}
where $\psi_{\bf{k}} \equiv \displaystyle{\begin{pmatrix}c_{\bf{k}\,\uparrow}\\c_{\,\bf{k}\,\downarrow}\end{pmatrix}}$
is the two-component electron spinor operator at wave-vector $\bf{k}$, $\mathrm{v}_{F}$ is the Fermi velocity, $\mu$ is the Fermi energy
(which lies above the Dirac point) and $\boldsymbol{\sigma} = (\sigma_1,\sigma_2)$ is the vector containing the first two Pauli matrices. The Dirac Hamiltonian (\ref{DH}) is diagonal in the helicity basis $\Psi_{{\bf k}}\,=\,\displaystyle{\begin{pmatrix} \gamma^{+}_{{\bf k}}\\\gamma^{-}_{{\bf k}}\end{pmatrix}}$:
\begin{align}
\Psi_{{\bf k}}\,&=\,U_{{\bf k}}\,\psi_{{\bf k}}\notag\\
U_{{\bf k}}\,&=\,\frac{1}{\sqrt{2}}\,\begin{pmatrix}\text{i}\,e\,^{\text{i}\,\varphi_{{\bf k}}}&1\\-\text{i}\,e\,^{\text{i}\,\varphi_{{\bf k}}}&1\end{pmatrix}
,\quad \varphi_{{\bf k}}\, \equiv\, \arctan{\left(\frac{k_{y}}{k_{x}}\right)}
\end{align}
yielding
\begin{equation}
\label{DHhb}
{\cal H}_{\rm {el}}\,=\,\sum_{\bf{k}}\,\sum_{\alpha = \pm}\,\xi^{\alpha}_{\bf{k}}\,(\gamma^{\alpha}_{\bf{k}})^{\dagger}\,\gamma^{\alpha}_{\bf{k}},\quad
\xi^{\alpha}_{\bf{k}} = \alpha\,\hbar\mathrm{v}_{F}\,|\bf{k}| - \mu 
\end{equation} The corresponding electron polarization function is defined in the RPA as
\begin{align}
\label{eq:RPA polarization}
\Pi({\bf q}, \mathrm{i}\omega_{n})\,=\,\frac{1}{\mathcal{A}}\,\frac{1}{\beta}\,\sum_{\mathrm{i}\Omega_{m}}\,\sum_{{\bf p}}\,\mathrm{Tr}&\left[
G^{(0)}_{\sigma,\sigma^\prime}({\bf p}+{\bf q},\mathrm{i}\Omega_{m}+\mathrm{i}\omega_{n})\right.\notag\\&\quad\left.G^{(0)}_{\sigma,\sigma^\prime}({\bf p},\mathrm{i}\Omega_{m})\right]
\end{align} 
where $\mathcal{A}$ is the surface area, and $\mathrm{Tr}$ acts on the spin degrees of freedom $\sigma, \sigma^\prime = \uparrow, \downarrow$. $G^{(0)}$ is the noninteracting electronic Matsubara Green's function with fermionic Matsubara frequencies $\Omega_m$
\begin{equation}
G^{(0)}_{\sigma,\sigma^\prime}({\bf p}, \mathrm{i}\Omega_{m} )\,=\,-\int^{\beta}_{0}\,\mathrm{d}\tau\,e\,^{\mathrm{i}\Omega_{m}\tau}\,\left<\,T_{\tau}\,
c_{{\bf p},\,\sigma}(\tau)\,c^{\dagger}_{{\bf p},\,\sigma^\prime}(0)\,\right>_{0}
\end{equation} where $\beta \equiv k_{B}\,T$ and $T_{\tau}$ the imaginary time-ordering operator. The RPA dielectric function is given by
\begin{equation}
\label{eq:dielectric function}
\varepsilon({\bf q}, \mathrm{i}\omega_{n})\,=\,1-v_c({\bf q}\,)\,\Pi({\bf q}, \mathrm{i}\omega_{n})
\end{equation}
where $v_c({\bf q}\,) = \frac{e^2}{2\varepsilon_{0}|{\bf q}|}$ is the two-dimensional Fourier transform of the electron-electron Coulomb interaction potential. 

We consider an interaction between the electron density and the displacement $\mathbf{u}_{j}$ of the $j^{\rm th}$ ion about its in-plane equilibrium position $\mathbf{R}^{(0)}_{j}$. The displacement $\mathbf{u}_j$ has both in-plane and out-of-plane components. The e-p interaction can be generically written as

\begin{equation}
{\cal H}_\text{el-ph}\,=\,\int\;d^2{\bf r}\;\rho_\text{el}({\bf r})\;\sum_{j=1}^N\;\boldsymbol{\eta}\left(\,{\bf r}-{\bf R}^{(0)}_{j}\right)\,\boldsymbol{\cdot}\mathbf{u}_j
\label{hep}
\end{equation}
where $\rho_{\rm{el}}({\bf r}) = \displaystyle{\sum_{\sigma = \uparrow,\downarrow}}\,c^{\dagger}_{\sigma}({\bf r})c_{\sigma}({\bf r})$ is the electron surface density operator and $\boldsymbol{\eta}(\,{\bf r}-{\bf R}^{(0)}_{j}\,)$ is a position dependent vector function (with units of energy per length) characterizing the e-p coupling. The quantities $\rho_{\rm{el}}$, $\boldsymbol{\eta}$ and $\mathbf{u}_j$ are then expanded in reciprocal-space as
\begin{align*}\boldsymbol{\eta}\left(\,{\bf r}-{\bf R}^{(0)}_{j}\right)\,&=\,\frac{1}{\mathcal{A}}\,\sum_{{\bf q}}\,\boldsymbol{\eta}_{{\bf q}}\;e\,^{\mathrm{i}{\bf q}\cdot({\bf r}-{\bf R}^{(0)}_{j})}\;\\[10pt]
\rho_{\rm{el}}({\bf r})\,& =\, \sum_{\sigma = \uparrow,\downarrow}\,c^{\dagger}_{\sigma}({\bf r})\,c_{\sigma}({\bf r})\\&=\, \frac{1}{\mathcal{A}}\,\sum_{\sigma = \uparrow,\downarrow}\,\sum_{\bf q}\;e\,^{-\mathrm{i}{\bf q}\cdot{\bf r}}\;\sum_{\bf k}\;c^{\dagger}_{{\bf k}+{\bf q},\,\sigma}\,c_{{\bf k},\,\sigma}\\\mathbf{u}_j\,&
=\,\frac{1}{\sqrt{N}}\,\sum_{{\bf q},\nu}\,\sqrt{\frac{\hbar}{2M \omega^{(0)}_{{\bf{q}},\nu}}}\,e\,^{\mathrm{i}{\bf q}\cdot{\bf R}^{(0)}_{j}}\,
(b_{{\bf q},\nu}+b^{\dagger}_{-{\bf q},\nu})\;\hat{\bf e}_{\nu}(\mathbf q)
\end{align*} where $N$ is the number of surface primitive cells and $\hat{\bf e}_{\nu}(\mathbf q)$ is the polarization vector. Substitution in (\ref{hep}) leads to the e-p interaction Hamiltonian 
\begin{equation}{\cal H}_\text{el-ph}\,=\,\frac{1}{\sqrt{\mathcal{A}}}\,\sum_{\sigma = \uparrow,\downarrow}\,\sum_{{\bf k}}\sum_{{\bf q},\nu}\,{\rm g}_{{\bf q},\nu}\,c^{\dagger}_{{\bf k}+{\bf q},\sigma}\,c_{{\bf k},\sigma}\;\left(b_{{\bf q},\nu}+b^{\dagger}_{-{\bf q},\nu}\right)\end{equation}
with the e-p coupling
$${\rm g}_{{\bf q},\nu}\, =\, \sqrt{\frac{N\hbar}{2 M \mathcal{A}\, \omega^{(0)}_{{\bf q},\nu}}}\;\boldsymbol{\eta}_{\mathbf q} \boldsymbol{\cdot}\hat{\bf e}_{\nu}(\mathbf q)\ \equiv\ \sqrt{\frac{N\hbar}{2 M \mathcal{A}\, \omega^{(0)}_{{\bf q},\nu}}}\;\eta_{{\mathbf q},\nu}$$
The e-p interactions renormalize the Matsubara phonon propagator, yielding  
\begin{align}\mathcal{D}_{\nu}({\bf q}, \mathrm{i}\omega_{n} )\,&=\,\frac{\mathcal{D}_{\nu}^{(0)}({\bf q}, \mathrm{i}\omega_{n})}
{1-\mathcal{D}_{\nu}^{(0)}({\bf q}, \mathrm{i}\omega_{n})\,|\text{g}_{{\bf q},\nu}|^2\,\frac{\Pi({\bf q}, \mathrm{i}\omega_{n})}{\varepsilon({\bf q}, \mathrm{i}\omega_{n})}}\notag\\&=\,\frac{2(\hbar\omega^{(0)}_{{\bf q},\nu})}{(\mathrm{i}\omega_{n})^2 - (\hbar\omega^{(0)}_{{\bf q},\nu})^2-2(\hbar\omega^{(0)}_{{\bf q},\nu})\tilde{\Pi}}\end{align} with $$\tilde{\Pi}=|\text{g}_{{\bf q},\nu}|^2\,\frac{\Pi({\bf q}, \mathrm{i}\omega_{n})}{\varepsilon({\bf q}, \mathrm{i}\omega_{n})}$$ being the phonon self-energy.
 

After performing the analytic continuation\hfill\break  $\mathrm{i}\omega_{n} \rightarrow \omega + \mathrm{i}0^+$, we obtain the renormalized frequencies as the real part of the poles of ${\cal D}_\nu({\bf q},\omega)$ 

\begin{equation}
(\hbar\omega_{{\bf q},\nu})^2\, =\, (\hbar\omega^{(0)}_{{\bf q},\nu})^2 +2(\hbar\omega^{(0)}_{{\bf q},\nu})\; \text{Re} \left[\tilde{\Pi}({\bf q}, \omega_{{\bf q},\nu})\right]\label{realpi}
\end{equation}
\text{Re}\,$\left[\tilde{\Pi}({\bf q}, \omega_{{\bf q},\nu})\right]$ is then adjusted to reproduce the measured phonon dispersion. It depends on two parameters, namely, the two components of the coupling function $\boldsymbol{\eta}=\left(\eta_\perp \ \mbox{and} \ \eta_{\|}\right)$ which lie in the sagittal-plane with
\begin{figure}
\begin{center}$
\begin{array}{cc}
\includegraphics[scale=0.30]{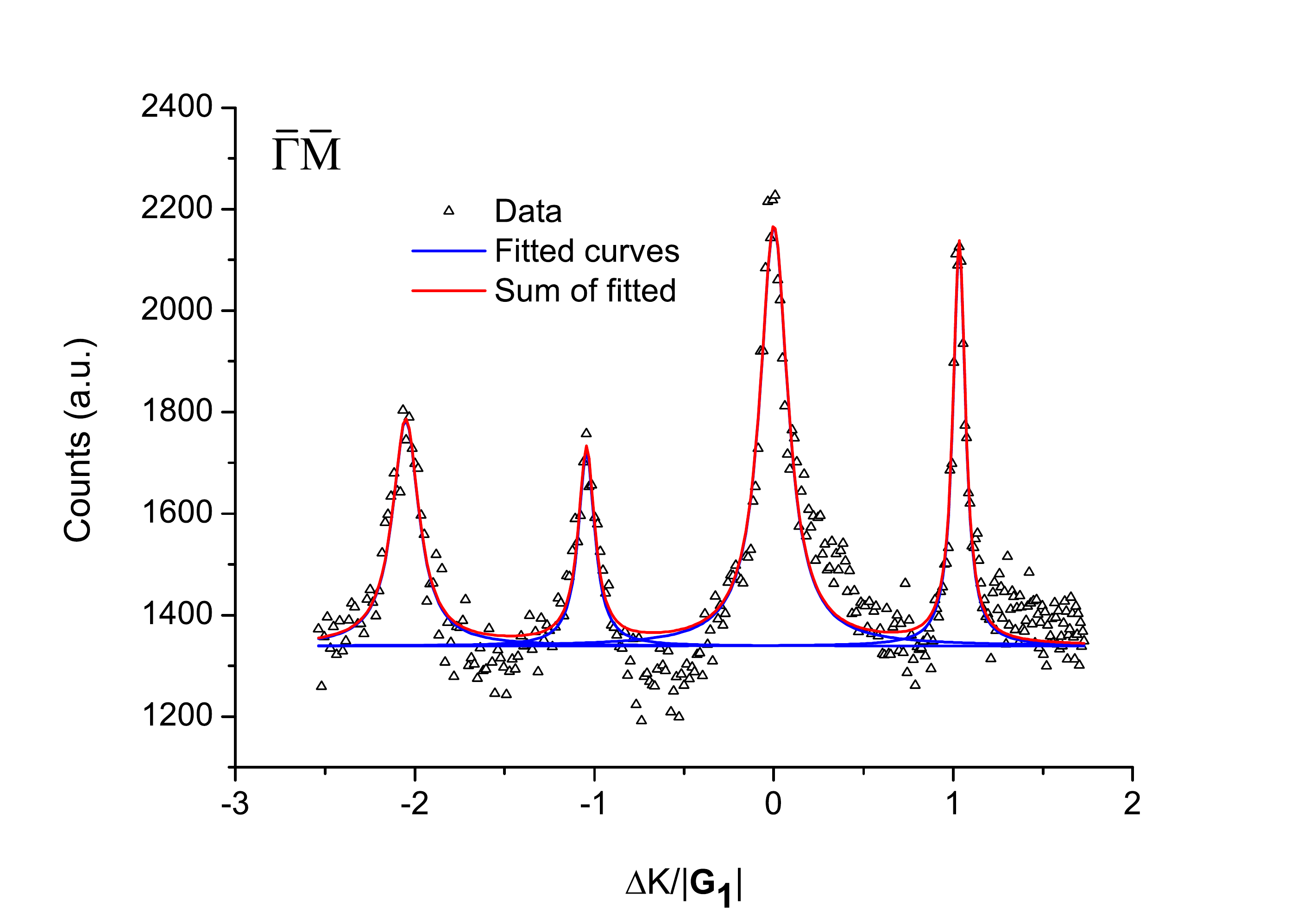} \\
\includegraphics[scale=0.30]{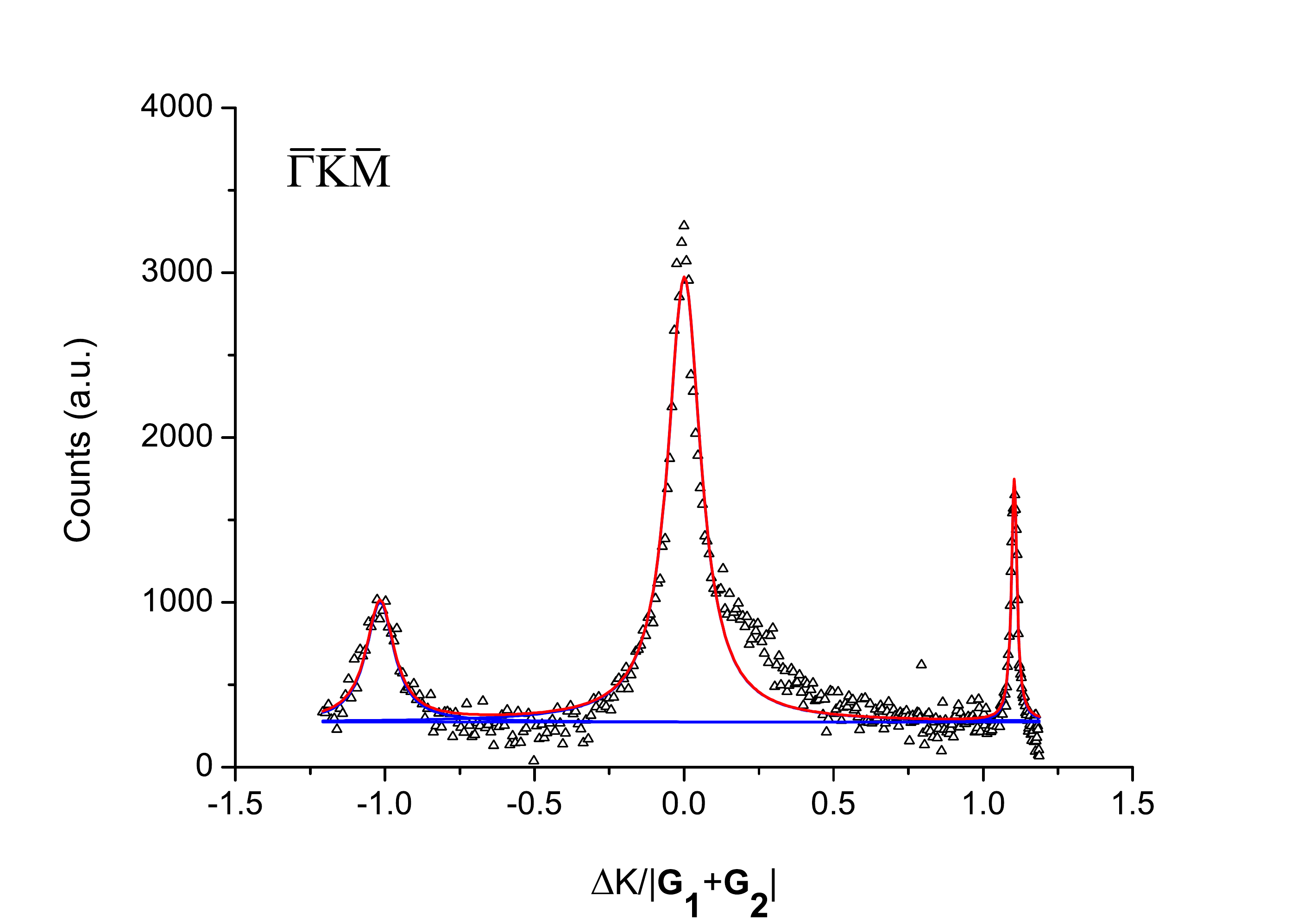}
\end{array}$
\end{center}
\caption{Diffraction patterns indicating the two high-symmetry directions $\bar\Gamma \bar{\mbox{M}}$ and $\bar\Gamma\bar{\mbox{K}}\bar{\mbox{M}}$ on the surface of Bi$_2$Te$_3$. The horizontal axis depicts momentum transfer as a fraction of the pertinent lattice vector.\label{diffraction} }
\end{figure}
\noindent directions normal and parallel to the wave-vector ${\bf q}$. A detailed definition of these couplings is given in Ref 7. The values of the bare phonon frequency $\omega^{(0)}$ and $k_F$ are extracted from experimental results. The former is identified as the experimental value of $\omega({\bf q}=0)=1.4$ THz, where the DFQ response vanishes, while $2k_F$ was set as the wave-vector where the V-shaped Kohn anomaly occurs\footnote{The similarity between the current results and those obtained for Bi$_2$Se$_3$ suggests that the termination of the Kohn anomaly is an indicator of suppressed DFQ screening and hence is a bona fide measure of $2k_F$.}.

After fitting $\text{Re} \left[\tilde{\Pi}({\bf q}, \omega_{{\bf q},\nu})\right]$ to the experimental dispersion curve, the corresponding $\text{Im} \left[\tilde{\Pi}({\bf q}, \omega_{{\bf q},\nu})\right]$ is obtained by a Kramers-Kronig transformation
\begin{equation}
\text{Im}\left[\tilde{\Pi}({\bf q}, \omega_{\bf q})\right] = \frac{2}{\pi}\int_0^\infty \frac{\omega_{\bf q}}{\omega_{\bf q}^2 - \omega_{\bf q}^{\prime2}} \,\text{Re}\left[\tilde{\Pi}({\bf q}, \omega_{\bf q}^\prime)\right] \ d\omega_{\bf q}^\prime
\label{KK}
\end{equation} 
Finally, the e-p coupling function is obtained from the relation\cite{Allen1,Allen2,Grimvall}
\begin{equation}
\lambda_\nu({\bf q}) = -\frac{\mbox{Im}[\tilde{\Pi}({\bf q},\omega_{{\bf q},\nu})]}{\pi {\cal N}(E_F)(\hbar\omega_{{\bf q},\nu})^2}
\label{lamq}
\end{equation}
where ${\cal N}(E_F)$ is the density of electronic states at the Fermi surface. 

\section{\label{sec:results}Experimental Results and Discussion}
\begin{figure}$
\begin{array}{cc}
\includegraphics[scale=0.17]{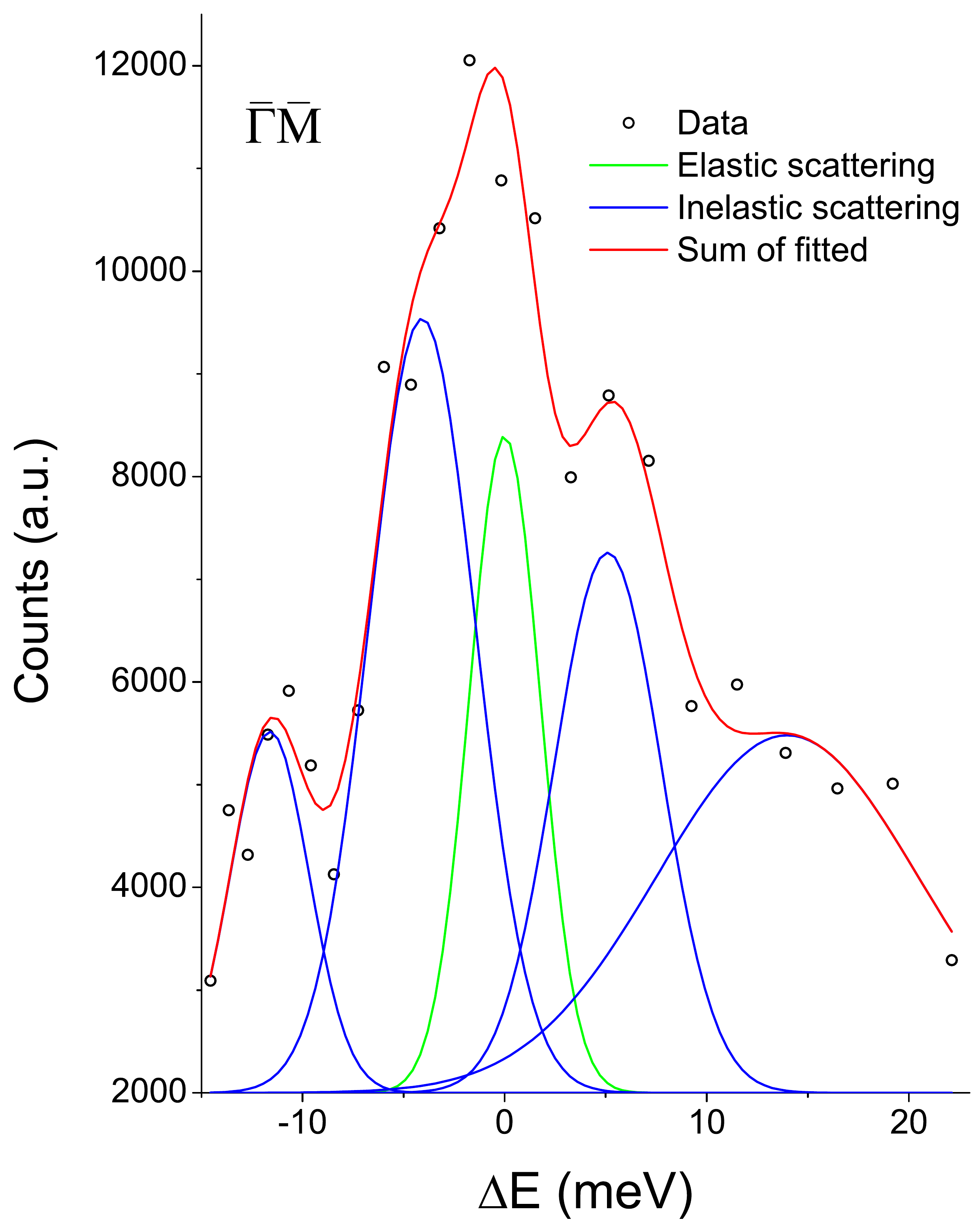} 
\includegraphics[scale=0.17]{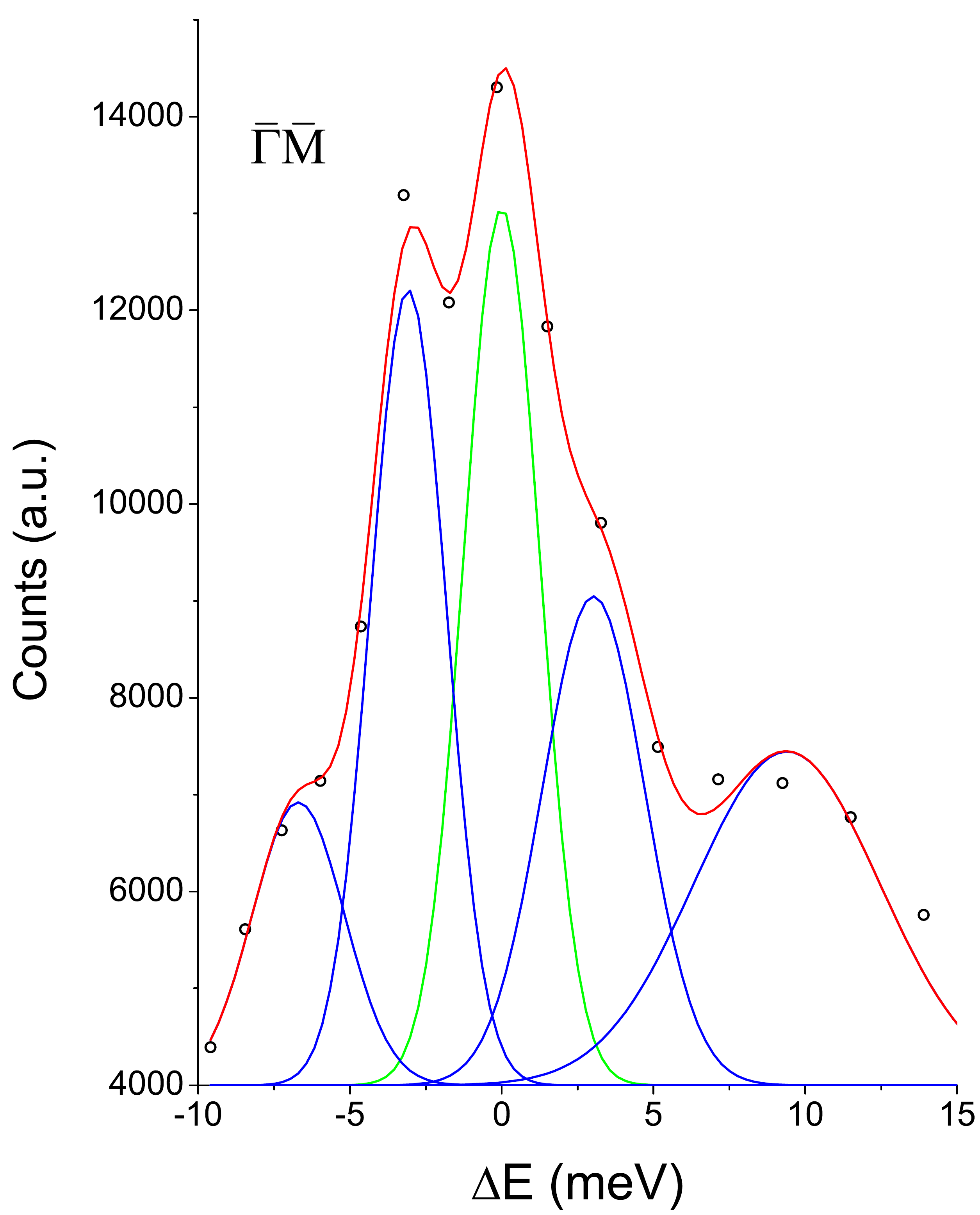}\\ 
\\
\includegraphics[scale=0.20]{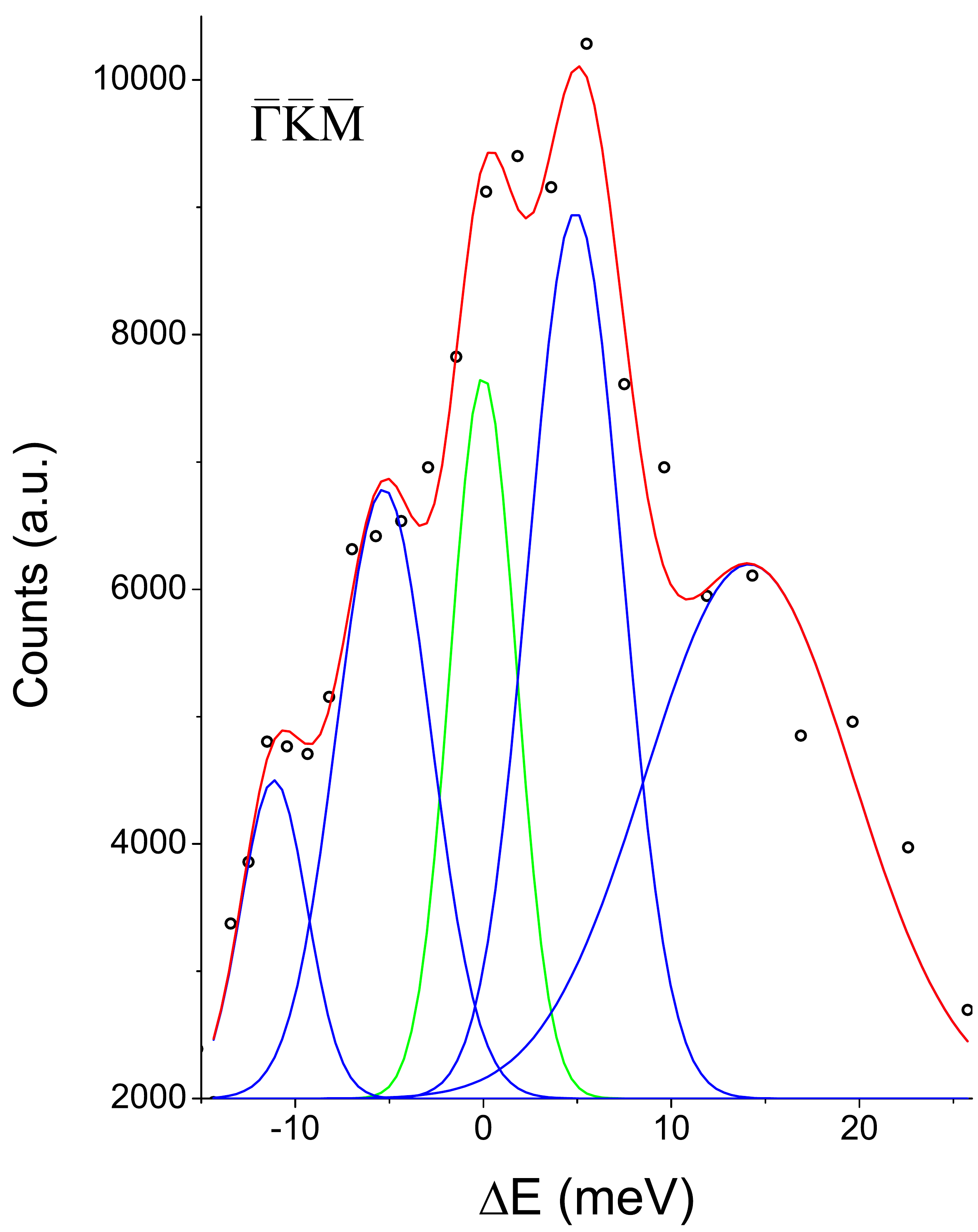}
\end{array}$
\caption{Time of flights scans for the inelastic HASS measurements. The abscissa has been converted to energy difference for clarity. Phonon creation and annihilation events are manifest as blue inelastic peaks. \label{TOF}}
\end{figure}

We begin by presenting in figure \ref{diffraction} typical diffraction patterns for both the $\bar\Gamma\bar{\mbox{M}}$ and $\bar\Gamma\bar{\mbox{K}}\bar{\mbox{M}}$ directions. The data shows good agreement with the values of the nominal lattice vectors. As was stated earlier, measurements of the phonon dispersion are obtained using inelastic HASS measurements and TOF techniques. Typical TOF scans collected along different high-symmetry directions are shown in figure \ref{TOF}.

\begin{figure*}
\includegraphics[scale=.7]{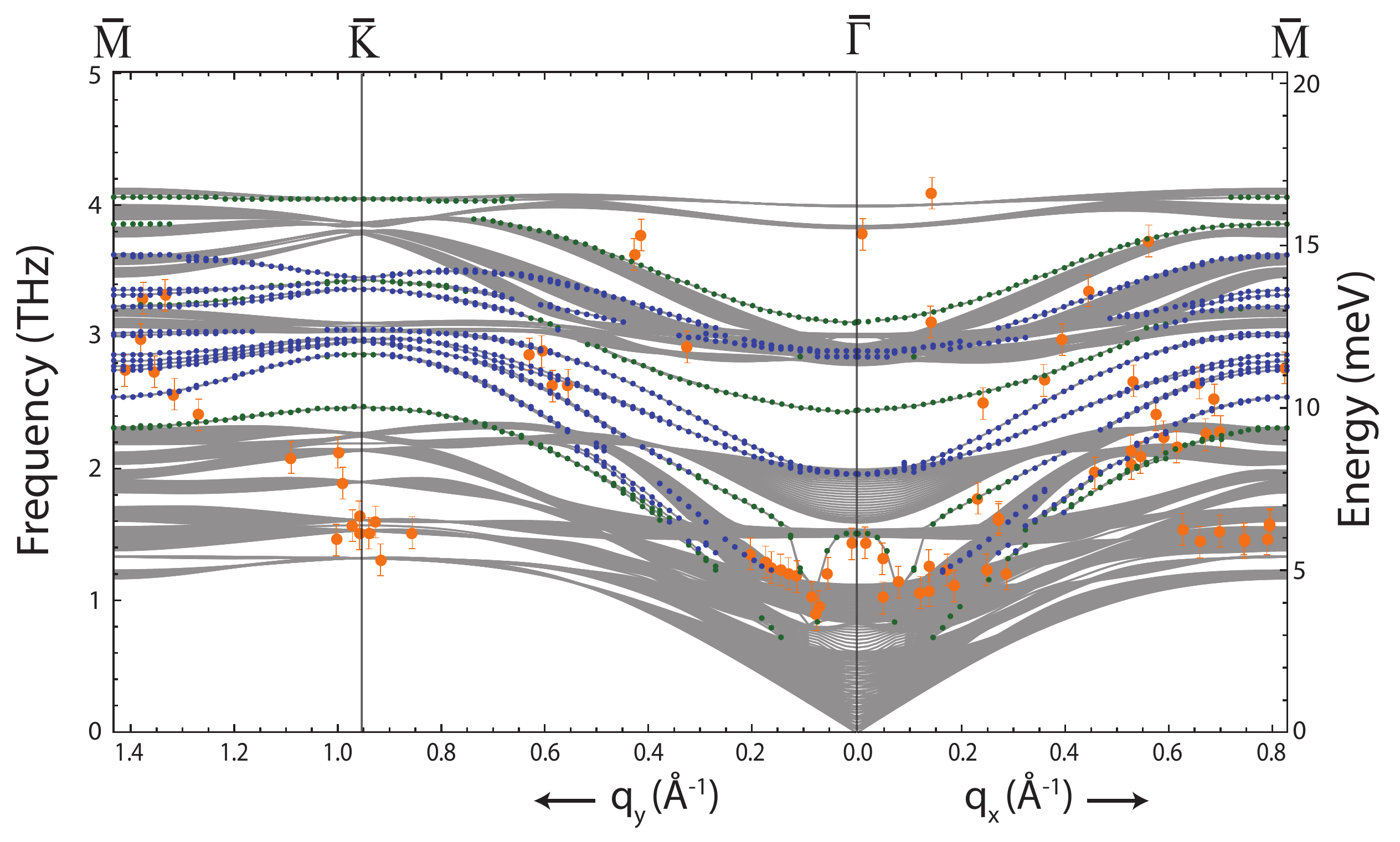}
\caption{Phonon dispersion along high-symmetry directions $\bar\Gamma \bar{\mbox{K}}\bar{\mbox{M}}$ and $\bar\Gamma \bar{\mbox{M}}$ for Bi$_2$Te$_3$. Gray areas represent the projection of bulk phonon bands onto the surface Brillouin zone whereas the dotted lines indicate surface phonon modes with at least 30\% of the oscillator strength concentrated within the first three layers of the 30 QL slab surface regions. Green and blue dots indicate modes polarized perpendicular and parallel to the surface plane, respectively. The TOF measurements are depicted as orange dots with error bars.\label{dispersion}}
\end{figure*}

\begin{figure}
\includegraphics[scale=0.32]{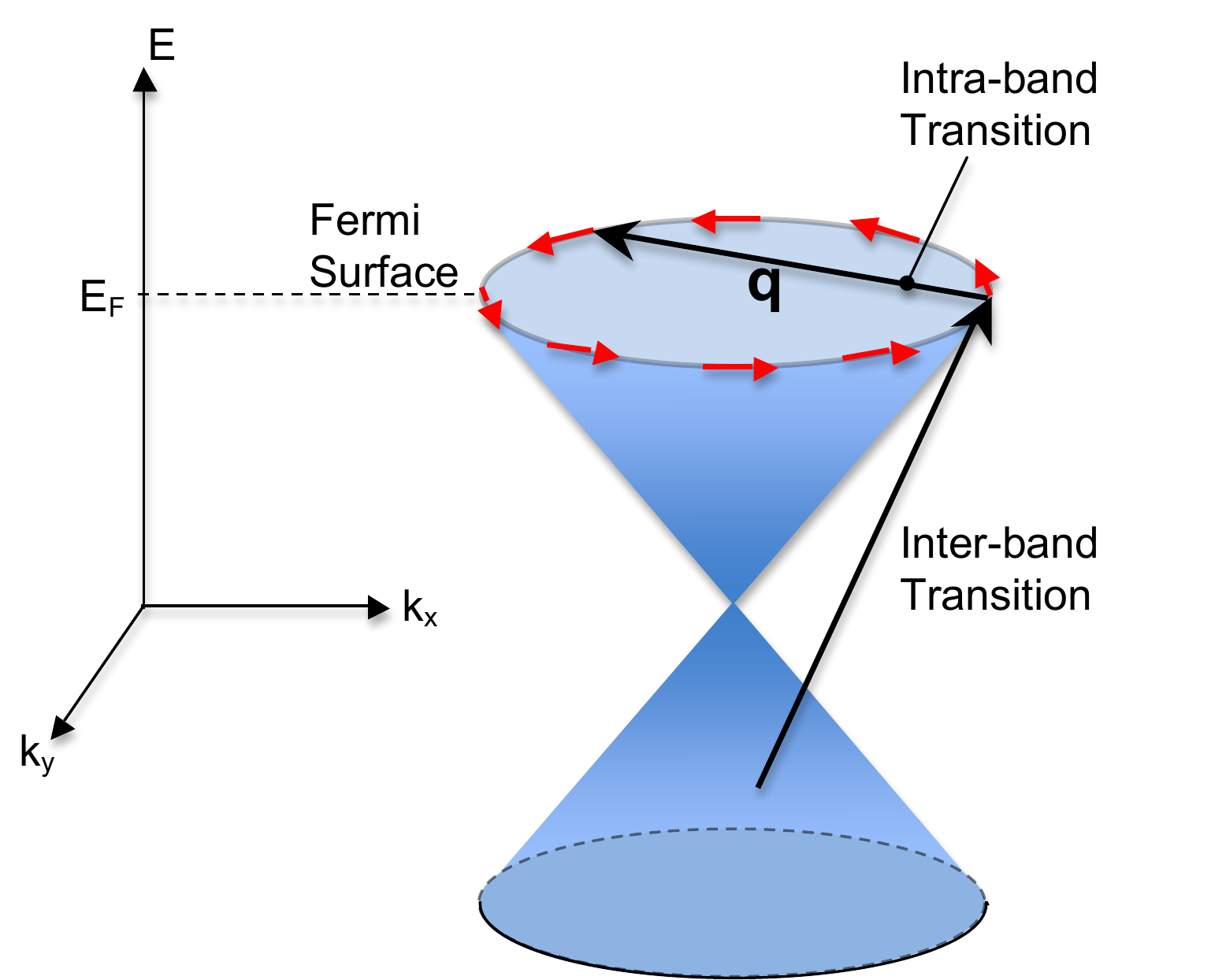}
\caption{Diagram showing the dispersion of the surface DFQ excitations. The spin chirality of the Fermi surface is depicted by the red arrows. The screening effects of the DFQs can be understood in terms of the scattering of electrons in the Dirac cone with the momentum transfer supplied by a phonon. We identify two distinct types of transitions: intra-band and inter-band, although the latter are suppressed by energetic considerations. Note that low-energy intra-band transitions are confined to a circle of diameter $2k_F$. \label{DiracCone}}
\end{figure}

The measured and computed dispersion curves for Bi$_2$Te$_3$ are presented along the two high-symmetry directions in figure \ref{dispersion}. 
Gray areas represent the projection of bulk phonon bands onto the surface Brillouin zone, whereas dotted lines indicate surface phonon modes with at least 30\% of the oscillator strength (determined by the square of the mode eigenvector) concentrated within the first three atomic layers of the slab surface regions. Green and blue dots signify $z$-polarized and $x/y$-polarized modes, respectively. There are two key features worth noting from the onset. First, we notice an optical surface phonon branch (hereon denoted $\nu$) originating at approximately 1.4 THz at the $\bar\Gamma$ point that disperses to lower energy with increasing wave-vector in both the $\bar\Gamma\bar{\mbox{M}}$ and  $\bar\Gamma\bar{\mbox{K}}$ directions. This trend terminates in a V-shaped minimum at  $q\approx$ 0.08\AA$^{-1}$ and $\omega\approx$ 1 THz, signifying a Kohn anomaly. This is consistent with our previous work\cite{zhu} on Bi$_2$Se$_3$ which exhibited a similar Kohn anomaly terminating at $q\approx$ 0.2\AA$^{-1}$. As before we attribute this phonon mode softening to an effective screening provided by scattering of the DFQs at the Fermi surface, which are schematically depicted in figure \ref{DiracCone}. 
Scattering events with a momentum transfer greater than the diameter of the Fermi surface require energy, and are therefore suppressed. This is manifest as the recovery of the $\nu$-branch dispersion after $q=2k_F\approx$ 0.08\AA$^{-1}$. Hence we have an operative DFQ screening for $q<2k_F$ which is attenuated for wave-vectors above this value\footnotemark[\value{footnote}].

We also note the absence of surface acoustic phonon modes in both the measured and computed dispersion, which is also consistent with our results in Bi$_2$Se$_3$. It appears that acoustic phonon modes with $q<2k_F$ are confined to the insulating bulk, unable to penetrate into the metallic surface layers due to mismatch in force-constants at the interface. However, for $q>2k_F$, these acoustic modes may resonate in the metallic film with their bulk counterparts as evidenced by the emergence of the z-polarized Rayleigh mode at $\omega\approx 0.8$ THz. We would like to point out that a recent study using density functional theory\cite{Huang} has also demonstrated the absence of long wavelength Rayleigh modes in the phonon dispersion. However, we should note that the authors' results are for Bi$_2$Te$_3$ thin films (2-3 QLs) so the connection to the current work may be tenuous. Indeed, the study also yields a value for the average e-p coupling constant that is significantly smaller than the result presented here. 

One may question why there are several low-energy measured events that do not overlap with any computed surface phonon modes. This may be attributable to large phonon density of states associated with a high concentration of flat, narrow projected bulk bands. It may be possible that we are sampling these bulk modes via surface resonances.  

\begin{figure}
\includegraphics[scale=0.15]{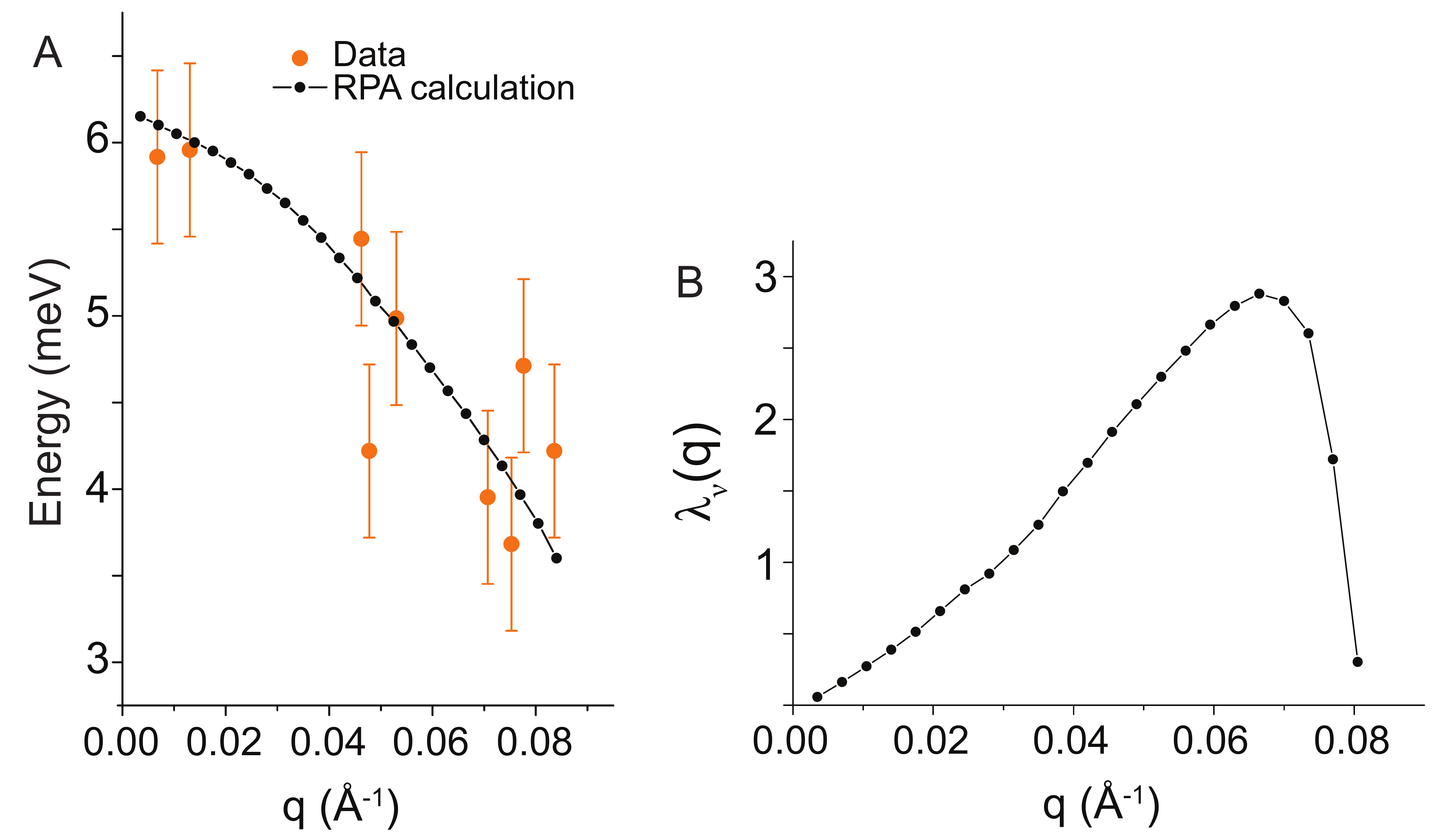} 
\caption{(A) Plot depicting the renormalization of the $\nu$-branch energy in the range $0<q<2k_F$. The black lines/dots indicate the renormalized phonon frequencies in the RPA calculation given by equation (\ref{realpi}), which are fit to experimental data (orange points). (B) Plot of the $\nu$-branch specific coupling function. The steady increase of  $\lambda_\nu({\bf q})$ with ${\bf q}$ is consistent with the softening of the $\nu$-branch phonon frequencies. The attenuation just before $q\approx2k_F$ is due to the fact that scattering of DFQs across the Fermi surface by phonons is not possible due to time reversal invariance.\label{RPAcalc}}
\end{figure}

In order to make quantitative statements about the magnitude of the e-p coupling on the surface of Bi$_2$Te$_3$ we fit the real part of the phonon self-energy to our experimental data as described in section \ref{sec:RPA}. Figure \ref{RPAcalc}A depicts the results where we overlapped the experimental data for the $\nu$-branch along both the $\bar\Gamma\bar{\mbox{M}}$ and  $\bar\Gamma\bar{\mbox{K}}$ directions. 
Next the imaginary part is obtained from the Kramers-Kronig transformation of the real part via equation (\ref{KK}). Finally we use equation (\ref{lamq}) to calculate the $\nu$-branch-specific e-p coupling function, which is plotted in figure \ref{RPAcalc}B. One will notice that $\lambda_\nu({\bf q})$ assumes quite large values with a maximum of $\approx 3$. Averaging over the points in figure \ref{RPAcalc}B yields $\langle\lambda_\nu\rangle = 1.44 $.

We note that, although large, this value for the coupling constant is compatible with recent results of high-resolution angle-resolved photoemission spectroscopy measurements of both Bi$_2$Se$_3$ and Bi$_2$Te$_3$ which give a value $\lambda\approx 3$\cite{Kondo}. However, the authors attribute this large value to contributions from e-p as well as electron-spin/plasmon interactions. It is possible that the latter could account for the residual contribution to the coupling constant. We also note that our calculated value of $\langle\lambda_\nu\rangle$ is significantly larger than a recent theoretical calculation\cite{Giraud} and an experimental study\cite{Chen2} using angle-resolved photoemission spectroscopy. 

\section{Conclusion}
We have measured the low-energy surface phonon dispersion of the strong TI Bi$_2$Te$_3$ using HASS techniques. A low-energy $z$-polarized optical surface phonon mode experiences strong downward renormalization in both the $\bar\Gamma\bar{\mbox{M}}$ and  $\bar\Gamma\bar{\mbox{K}}$ directions, signifying a Kohn anomaly terminating at approximately $2k_F$. Moreover, we find no trace of the ubiquitous Rayleigh surface mode for phonon wave-vectors below this value. Our measurements are substantiated by lattice dynamical calculations based upon the PCM, which capture both of these phenomena when interactions between surface PC are taken into account.

In addition, we have performed a calculation of the e-p coupling constant for the uniquely dispersive surface optical phonon branch. By fitting our experimental data to a phenomenological model based in the RPA we find $\langle\lambda_\nu\rangle$ = 1.44. This exceptionally large value is consistent with a recent experimental study using angle-resolved photoemission spectroscopy, but disagrees with some other theoretical and experimental claims. Thus a consensus about the strength of the e-p interaction in Bi$_2$Te$_3$ has yet to be reached. 

\section*{Acknowledgements}
The authors would like to thank L. Santos and C. Chamon for critical contributions to the theoretical calculation of the e-p coupling constant, and X. Zhu for valuable discussions relating to data acquisition and analysis. This work is supported by the U.S. Department of Energy under Grant No. DE-FG02-85ER45222. FCC acknowledges the support from the National Science Council of Taiwan under project No. NSC 99-2119-M-002-011-MY.       
\bibliographystyle{apsrev4-1}
\bibliography{refs}
\end{document}